\documentclass[pra,twocolumn,showpacs,preprintnumbers,amsmath,amssymb,superscriptaddress]{revtex4}

\usepackage{graphicx}
\usepackage{dcolumn}
\usepackage{bm}

\def \ed {\end{document}}
\def\Fbox#1{\vskip1ex\hbox to 8.5cm{\hfil\fboxsep0.3cm\fbox{%
  \parbox{8.0cm}{#1}}\hfil}\vskip1ex\noindent}  

\def\be{\begin{equation}}\def\ee{\end{equation}}
\def\bea{\begin{eqnarray}}\def\eea{\end{eqnarray}}
\def\bse{\begin{subequations}}\def\ese{\end{subequations}}
\newcommand{\BE}[1]{\begin{equation}\label{#1}}
\newcommand{\BEA}[1]{\begin{eqnarray}\label{#1}}
\newcommand{\BSE}[1]{\begin{subequations}\label{#1}}

\let \= \equiv \let\*\cdot \let\~\widetilde \let\^\widehat \let\-\overline

\def\<{\left\langle}    \def\>{\right\rangle}
\def\({\left(}          \def\){\right)}
 \def \[ {\left [} \def \] {\right ]}

\makeatletter
\AtBeginDocument{\@ifpackageloaded{natbib}{\ifNAT@numbers\if@filesw\immediate\write\@auxout{\string\global\string\NAT@numberstrue}\fi\fi}{}}
\makeatother
\begin{document}

\preprint{APS/123-QED}

\title{Spin-glass-like behavior in the spin turbulence of spinor Bose-Einstein condensates }

\author{Makoto Tsubota}
\affiliation{Department of Physics, Osaka City University, Sumiyoshi-ku, Osaka 558-8585, Japan}
\affiliation{The OCU Advanced Research Institute for Natural Science and Technology (OCARINA), Osaka City University, Sumiyoshi-ku, Osaka 558-8585, Japan}

\author{Yusuke Aoki}
\affiliation{Department of Physics, Osaka City University, Sumiyoshi-ku, Osaka 558-8585, Japan}

\author{Kazuya Fujimoto}
\affiliation{Department of Physics, Osaka City University, Sumiyoshi-ku, Osaka 558-8585, Japan}

\date{\today}

\begin{abstract}
We study numerically the spin turbulence (ST) in spin-1 ferromagnetic spinor Bose-Einstein condensates (BECs).
ST is characterized by a $-7/3$ power law in the spectrum of the spin-dependent interaction energy.
The direction of the spin density vector is spatially disordered but temporally frozen in ST, showing an analogy with the spin glass state.
Thus, we introduce the order parameter of spin glass into ST in spinor BECs.
When ST develops through some instability, the order parameter grows with a $-7/3$ power law, thus succeeding in describing  ST well.
\end{abstract}

\pacs{03.75.Mn, 03.75.Kk}

\maketitle
Turbulence has been a great mystery in nature \cite{Davidson} mainly because it is a complicated dynamical phenomenon with strong nonlinearity far from an equilibrium state.
When we consider such complicated phenomena, it is important to focus on  statistical laws.
The most important statistical law in turbulence is the Kolmogorov $-$5/3 law.
The Kolmogorov spectrum is confirmed  in fully developed turbulence in classical fluids, being the smoking gun of turbulence.

Apart from these studies in classical fluid dynamics, there have been studies on quantum fluids such as superfluid helium and atomic Bose-Einstein condensates (BECs).
A quantized vortex appears as a stable topological defect in quantum fluids.
Such vortices give rise to quantum turbulence (QT) \cite{Hal,Novel, SkrbekReview}.
Quantum turbulence has  long been studied in superfluid helium, whereas atomic BEC has only recently become an important  research area.

An important interest on QT is how to characterize the turbulent state.
Numerical simulation of the Gross-Pitaevskii (GP) model shows that the spectrum of imcompressible kinetic energy obeys the Kolmogorov $-$5/3 law in a uniform system \cite{Nore, KT05} and in a trapped system \cite{KT07}.
Confirmation of such an energy spectrum power law would certainly provide strong proof of turbulence.
However, there must be another better way to characterize QT.

Spinor BECs have spin degrees of freedom and exhibit phenomena characteristic of spin \cite{KU}, which can be another novel stage of turbulence in quantum fluids.
The hydrodynamics of spinor BECs has recently been studied by several authors \cite{Lamacraft,Barnett,KK, YukawaUeda12}.
In  previous studies, we found that hydrodynamic instability in spin-1 spinor BECs exhibits unique behavior owing to their spin degrees of freedom and 
form spin turbulence (ST) in which the spin density vector has various directions.
Although the spin-dependent interaction can be ferromagnetic or antiferromagnetic, we confine ourselves to the case of ferromagnetic interaction in this work.
The first study addressed the counterflow between the $m=\pm 1$ components in a uniform system, where $m$ is the magnetic quantum number.
Through the instability the counterflow leads to ST, in which the spectrum of the spin-dependent interaction energy obeys a $-7/3$ power law \cite{FT12a}; 
the $-7/3$ power law is understood by the scaling analysis of the time-development equation of the spin density vector.
Such a spin-disordered state was  experimentally created in a trapped system through the instability of the initial helical structure of spins \cite{Berkeley08}, which allowed  us to study ST numerically in a similar situation and confirm the $-7/3$ power law again \cite{FT12b}.
An oscillating magnetic field applied to a uniform ferromagnetic system can also create ST with a $-7/3$ power law \cite{AT13}.

These three works \cite{FT12a,FT12b,AT13} reveal the important characteristics of ST.
First, the $-7/3$ power law is robust independently of whether the system is uniform or trapped, or how the system is excited.
Second, observation of the spin motion in all three cases indicates that the spin density vectors become spatially random but temporally frozen, which reminds us of spin glass.
Spin glasses are magnetic systems in which the interactions between the magnetic moments are in conflict with each other \cite{RMP}.
Thus, these systems have no long-range order but exhibit a freezing transition to a state with a kind of order in which the spins are aligned in random directions.

In this paper, we introduce the order parameter of spin glass \cite{RMP,SK} to characterize ST in spinor BECs.
The spin-glass like behavior is shown clearly in the movies of ST in Supplemental Material (SM) \cite{SM}. 
Movie (a.mpg) shows how the spins become random in the early period $t/\tau = 60 \sim 360 $ through the counterflow instability, 
where $\tau$ is the characteristic time in the uniform system \cite{FT12a}. 
Movie (b.mpg) in the late period $t/\tau = 4000 \sim 4300 $ shows the behavior of spins after the -7/3  power law appears. 
We can find the clear difference between two movies; compared with movie (a.mpg), the random spins in movie (b.mpg) look to oscillate at each site rapidly with small amplitude.
This introduction of the order parameter of spin glass is so successful that it grows with the setup of the $-7/3$ power law.
This success paves the way for two innovative approaches.
One is to propose a useful order parameter in the turbulence of quantum fluids.
The other is to connect the study of spinor BECs with that of magnetism including spin glass.
In this paper, first we describe how the ST in spinor BECs behaves.
Second, the introduction of the spin glass order parameter is shown to successfully characterize the ST.

\begin{figure} [t]
\begin{center}
\includegraphics[width=65mm]{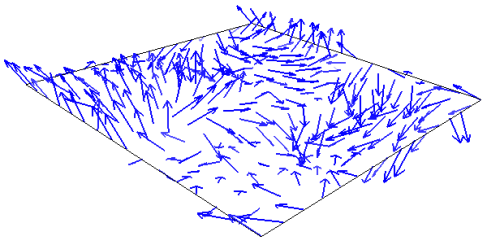}
\caption{(Color online) Distribution of spin density vectors in ST at $t/\tau = 4000$ obtained by the counterflow
instability with $V_{R}/c_{sound} = 0.78$, where $V_{R}$, $c_{sound}$, and $\tau$ are the relative velocity of counterflow, the sound
velocity, and the characteristic time in a uniform system \cite{FT12a}. The system size is $128 \xi \times 128 \xi$ with the coherence length $\xi$.
This numerical calculation starts from the initial state of the counterflow between the $m=\pm 1$ components, which induces the spin modulation,
finally leading to ST \cite{FT12a}. }
\label{fig:1}
\end{center}
\end{figure}

\begin{figure*} [htbp]
\begin{center}
\includegraphics[width=180mm]{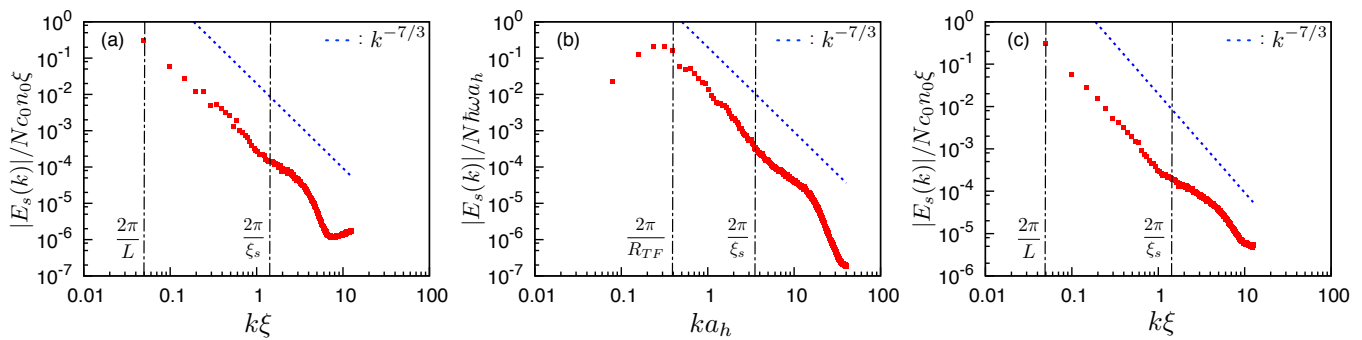}
\caption{(Color online) Spectrum of the spin-dependent interaction energy in three cases: (a) counterflow instability in a uniform system, (b) instability of the initial helical structure of the spin density vector in a trapped system, and (c) application of an oscillating magnetic field in a uniform ferromagnetic system.  Each spectrum is calculated at (a) $t/\tau = 4000$, (b) $t\omega = 200,$ and (c) $t/\tau = 9000$, when ST is fully developed. 
In the case of (c), the magnetic field is tuned off at $t / \tau = 5000$. 
The parameters $\omega$, $a_{h}$, $R_{TF}$, $n_{0}$, $\xi$, and $L$ are, respectively, the trapping frequency, the harmonic oscillator length, the Thomas-Fermi radius, the total density in the uniform system, the coherence length, and the system size. The details of these parameters are described in \cite{FT12a, FT12b, OMF2}, respectively. The blue broken lines show the $-7/3$ power law. The spectrum in all cases exhibits the $-7/3$ power law in the region [(a) $ 2 \pi / L < k <  2 \pi / \xi _{s} $ , (b) $2 \pi / R_{TF} < k < 2 \pi / \xi _{s}$, (c) $ 2 \pi / L < k < 2 \pi / \xi _{s} $] \cite{vor}, where $\xi _{s}$ is the spin coherence length.}
\label{fig:2}
\end{center}
\end{figure*}

We consider a spin-1 spinor BEC at zero temperature.
 The macroscopic wave functions $\psi _m$  ($m = 1,0,-1$) obey the GP equation \cite{Ohmi98, Ho98}
 \begin{eqnarray}
 i\hbar \frac{\partial}{\partial t} \psi _{m} &=&  (-\frac{\hbar ^2 }{2M} \nabla ^2 + V) \psi _{m} \nonumber \\
 &+&  \sum _{n=-1} ^{1} [- g \mu _{B}(\bm{B} \cdot \hat{\bm{S}})_{mn} + q(\bm{B} \cdot \hat{\bm{S}})^{2}_{mn} ]\psi _{n}  \nonumber \\
 &+& c_{0} n \psi _{m} +  c_{1} \sum _{n=-1} ^{1} \bm{s} \cdot \hat{\bm{S}} _{mn} \psi _{n}.  \label{2dGP}
 \end{eqnarray}
Here, $V$ and $\bm{B}$ are the trapping potential and magnetic field. The parameters $M$, $g$, $\mu _{B}$, and $q$ are the mass of a particle,
the Land$\rm \acute{e}$ $g$ factor, the Bohr magneton, and a coefficient of the quadratic Zeeman effect, respectively.
 The total density $n$ and the spin density vector $s_{i}$ ($i = x, y, z$ ) are given by $n =  \sum _{m=-1} ^{1}|\psi _m|^2$ and  $s_{i} = \sum _{m,n = -1}^{1} \psi _{m}^{*} (\hat{S}_{i})_{mn} \psi _{n}$ with
 the spin-1 matrices $(\hat{S}_{i})_{mn}$. The parameters $c_{0}$ and $c_{1}$ are the coefficients of the spin-independent and spin-dependent interactions.
We focus on the spin-dependent interaction energy $E_{spin} = \frac{c_{1}}{2} \int \bm{s} ^{2} d \bm{r}$, whose coefficient $c_{1}$ determines whether the system is ferromagnetic ($c_{1} < 0$) or antiferromagnetic ($c_{1} > 0$).
We are interested in the ferromagnetic case where ST clearly exhibits the $-7/3$ power law.

In this paper, we study the two-dimensional ST obtained by (i) the counterflow instability \cite{FT12a}, (ii) the instability of spin helical structure \cite{FT12b}, and (iii) an oscillating magnetic field \cite{AT13}. All parameters
of the numerical calculation of (i) and  (ii) are the same as in the previous studies.
In the case of (iii), we use a method different from the previous study \cite{AT13}; the magnetic field is turned off after the formation of ST \cite{OMF1}. The parameters are different from the previous study too \cite{OMF2}.

Figure 1 shows a typical case of ST developing from the instability of the counterflow between the $m=\pm 1$ components \cite{FT12a}.
As ST develops, the spectrum of the spin-dependent interaction energy begins to exhibit the $-7/3$ power law.
Figure 2 shows that the $-7/3$ power law appears in all three cases we have  studied:
counterflow instability in a uniform system [Fig. 2(a)] \cite{FT12a},
instability of the initial helical structure of the spin density vector in a trapped system [Fig. 2(b)] \cite{FT12b,trap}, and
application of an oscillating magnetic field in a uniform ferromagnetic system [Fig. 2(c)] \cite{AT13}.
All cases show that the spin density vector is spatially random but temporally frozen.
Such a behavior of spins reminds us of the analogy with spin glass, which invites us to introduce the order parameter of spin glass. 

The order parameter of spin glass is ordinarily introduced as  follows:
An equilibrium system is supposed to consist of $N$ lattice sites, each site $i$ having spin $\bm{S}_i$ ($i=1,2, \cdot \cdot \cdot ,N$).
The time average $\langle \bm{S}_i (t)\rangle$ and the space average $[\bm{S}_i (t)]$ are defined as
\begin{equation}
\langle \bm{S}_i (t)\rangle=\lim_{T \rightarrow \infty} \frac1T \int_0^T \bm{S}_i (t)dt, \;[\bm{S}_i (t)]=\frac1N \sum\limits_i  \bm{S}_i (t).  \label{average}
\end{equation}
 Then it is possible to introduce two order parameters, namely, the magnetization $\bm{M}=[\langle \bm{S}_i (t)\rangle]$ and $q=[\langle \bm{S}_i (t)\rangle^2]$ \cite{RMP,SK}.
If the system is paramagnetic, both $\bm{M}$ and $q$ vanish.
A ferromagnetic order gives nonzero $\bm{M}$ and $q$.
When $\bm{M}=0$ but $q$ is nonzero, the state is called spin glass \cite{SK}, and the directions of spin are spatially random but temporally frozen.

To apply the order parameter to our case, we have to consider two things.
First, we address the spin density vector $\bm{s} (\bm{r},t)$ instead of $\bm{S}_i (t)$.
Our spinor BECs are usually trapped so that the amplitude of the spin density vector is not uniform.
To extract this effect and focus on the direction of the spin density vector, we define the order parameter with the unit vector $\hat{\bm{s}}(\bm{r},t)=\bm{s}(\bm{r},t)/|\bm{s}(\bm{r},t)|$.
The definition of the space average of Eq. (2) is replaced by
\begin{equation} [\hat{\bm{s}} (\bm{r},t)]=\frac1A \int_A \hat{\bm{s}} (\bm{r},t)d\bm{r} \label{spaceaverage}\end{equation} with  the system area $A$.
Second, the system starts from some initial state to develop  ST, which should be characterized by the time-dependent order parameter.
Thus, we introduce the time average of $\hat{\bm{s}} (\bm{r},t)$ during the period $[t, t+T]$ as
\begin{equation}
\langle \hat{\bm{s}} (\bm{r},t)\rangle_T= \frac1T \int_t^{t+T}\hat{ \bm{s}} (\bm{r},t_1)dt_1
\end{equation}
and define the time-dependent order parameter
\begin{equation}
q(t)=[\langle\hat{ \bm{s}} (\bm{r},t)\rangle_T^2].
\end{equation}

When we calculate $q(t)$, we should be careful with how to take $T$.
Generally, a longer $T$ is desirable, but in reality, we should estimate $q(t)$ with some finite $T$.
The criterion for the appropriate value of $T$ would be that $T$ should be longer than the characteristic time of the system.
In this system, the velocity characteristic of spin is given by $c _{s} = \sqrt{|c_{1}|n/M}$ \cite{cs}.
Then, a typical characteristic time is the system size $L$ divided by the velocity $c_s$.
We will show in the following that this time $L/c_{s}$ is so long that the system in the initial state becomes sufficiently disturbed by the time.
Thus, we take $T$ comparable to $L/c_s$.
In ST, the order parameter $q(t)$ decreases with $T$, but the dependence is weak.
If the spin density vector is completely frozen in ST, $q(t)$ should be unity.

\begin{figure*} [htbp]
\begin{center}
\includegraphics[width=185mm]{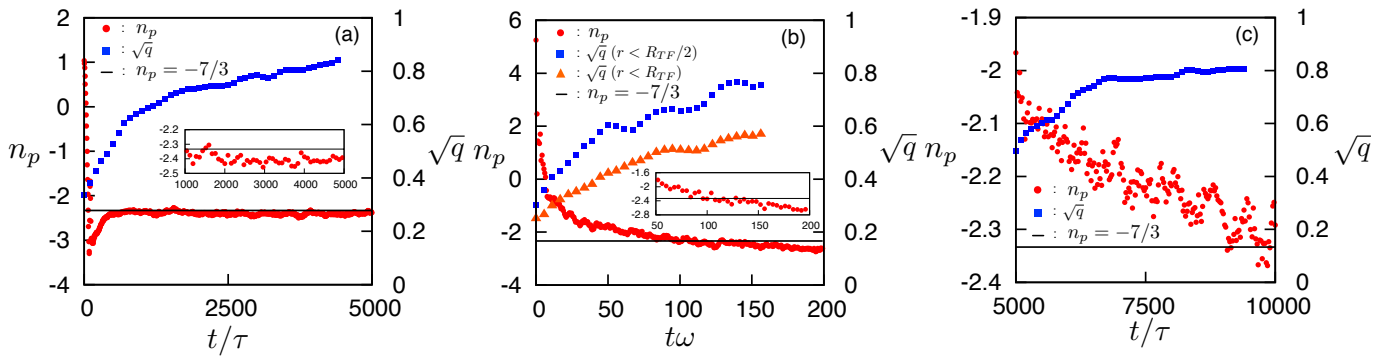}
\caption{(Color online) Time development of the exponent $n_{p}$ and the spin glass order parameter $ \sqrt{q(t)} $ for (a) the counterflow instability in a uniform system, (b) the instability of the initial helical structure of the spin density vector in a trapped system, and (c) the application of the oscillating magnetic field in a uniform ferromagnetic system. In (b), the squares and triangles show that the calculation of $ \sqrt{q} $ is performed in the regions $r < {R_{TF}/2}$ and $r < {R_{TF}}$, respectively. The exponents $n_{p}$ in (a) $\sim$ (c) are evaluated by the least-squares method in the region [(a) $ 4 \pi / L < k <  2 \pi / \xi _{s} $ , (b) $2 \pi / R_{TF} < k < 2 \pi / \xi _{s}$, (c) $ 4 \pi / L < k < 2 \pi / \xi _{s} $]. Then, in the uniform systems of (a) and (b), we omit the point $2\pi /L$, where the spectrum can be affected by the finite size effect. The insets of (a) and (b) are the enlarged graphs of $n_{p}$. }
\label{fig:3}
\end{center}
\end{figure*}

We show how $q(t)$ grows toward ST with a $-7/3$ power law of the spin-dependent interaction energy.
Figure 3 shows the time dependence of the power exponent of the spectrum of the spin-dependent interaction energy obtained by the least-squares method and $q(t)$ for the cases of Figs. 2(a), (b) and (c).
Figures 3(a) and 3(b) show that $q(t)$ increases obviously as the exponent $n_{p}$ approaches $-7/3$.
Furthermore, the magnetization $|\bm{m}(t)|= |[\langle\hat{\bm{s}} (\bm{r},t)\rangle _{T}]|$ is much smaller than $\sqrt{q(t)}$
because $[{\bm{s}} (\bm{r},t)]$ is conserved to keep vanishing under Eq. (1) without the magnetic field.
On the other hand, in the case of Fig. 3 (c), the state at $t/\tau =5000 $ is much disturbed \cite{OMF1}, so that the power exponent $n_{p}$ is already close to $-7/3$. As the time passes, the exponent approaches to $-7/3$ and $q(t)$ grows too. 
Also, the magnetization $|\bm{m}(t)|$ is almost zero at $t/\tau=5000$.
Thus, the two order parameters $q(t)$ and $|\bm{m}(t)|$ are found to be effective for describing the ST in the three cases.

Only the case (b) is affected by the inhomogeneity.
Because the system is trapped, the amplitude of the spin density vector is reduced going from the center to the boundary.
The low condensate density near the boundary induces a large fluctuation of the condensate phase, thus causing  $\bm{s} (\bm{r},t)$ to fluctuate temporally.
As a result, the space average of Eq. (\ref{spaceaverage}) depends on the area $A$ of the integral.
Figure 3(b) shows  two cases for which the radius of $A$ is $R_{TF}/2$ and $R_{TF}$ with the Thomas-Fermi radius being $R_{TF}$.
The larger value of $q(t)$ for $R_{TF}/2$ means that the central spins are more likely to be frozen.

We note that the time $T = L/c_{s}$ is so long that the system becomes enough disturbed, which is found by the time  development of $n_{p}$.
The time $T$ is (a) $ 572 \tau $, (b) $ 40 / \omega, $ and (c) $ 572 \tau $, respectively. We find that $n_{p}$ rapidly changes in the period $0 < t < T$, which means 
that the distribution of the spin density vector in the wavenumber space changes rapidly too. 
Thus, in this period, the spin density vector temporally can point in various directions. 
Therefore, the time $L/c_{s}$ is appropriate for the calculation of the spin glass order parameter $q(t)$.

We do not know the origin of the spin-glass-like behavior currently. 
Of course, our system of spinor BECs is much different from a magnetic system yielding spin glass, 
and the spin turbulence is not spin glass. 
One possible mechanism causing the spin-glass-like behavior may be related with the growth of large-scale spin structures, 
which is reported for two-component BECs\cite{Karl13} and spin-1 ferromagnetic spinor BECs under magnetic fields \cite{KK13}. 
We will report this issue shortly.  

In summary, we have studied numerically ST in spin-1 spinor BECs.
The spectrum of the spin-dependent interaction energy is found to exhibit a $-7/3$ power law independently of the details of the system or how  the ST is created.
The direction of the spin density vector is spatially disordered but temporally frozen in ST, which shows an analogy with the spin glass state.
Thus we introduced the order parameter of the spin glass into ST in spinor BECs.
The order parameter $q(t)$ grows with a $-7/3$ power law,  well describing  ST.
These behaviors should be accessible experimentally. Some unsolved problems of ST are described in SM \cite{SM2}.

 We thank Shin-ichi Sasa for useful discussions.


\begin{thebibliography}{99}
\bibitem{OMF1} We apply an oscillating magnetic field $\bm{B} = \bm{B}_{0} {\rm sin}(\omega _{M}t) $ along the $x$ direction with the frequency $\omega _{M} = 2 \pi c_{0}n_{0}/100 \hbar$. After ST appears, we turn off the magnetic field at $t/\tau=5000$ and investigate the time-development of the system.
\bibitem{OMF2}  We transform Eq. (1) to the nondimensional equation by using the characteristic length $\hbar / \sqrt{2Mc_{0}n_{0}}$ and time $\hbar /c_{0}n_{0}$. Here, $n_{0} = N/L_{x}L_{y}$ is the initial total density, where $N$, $L_{x}$, and $L_{y}$ are the total particle number and the system size in the $x$ and $y$ directions, respectively. The nondimensional system size $ \tilde{L}_{x} \times  \tilde{L}_{y}$ is $128  \times 128,$ which is discretized into $512 \times 512$ bins. In our numerical calculation, the parameters $ \tilde{L}_{x}\tilde{L}_{y}$, $c_{1} \tilde{L}_{x} \tilde{L}_{y}/c_{0}$, $ g \mu_{B}|\bm{B}_{0}|/c_{0}n_{0}, $ and $q |\bm{B}_{0}|^{2}/c_{0}n_{0}$ are $128^{2}$, $128^{2}/20$, $0.1$,  and $0.1$, which are the coefficients of the spin-independent interaction, the spin-dependent interaction, the linear Zeeman,  and the quadratic Zeeman terms in the nondimensional form of Eq. (1), respectively.
\bibitem{trap} In the case of Fig. 2(b), the total density is not uniform, which leads to the deviation from the -7/3 power because the uniformity of the total density is assumed in the derivation of the $-7/3$ power law \cite{FT12a}.
\bibitem{vor} In the wave number region higher than $2 \pi / \xi_{s}$, the spectrum may reflect the structure of spin vortices and domain walls. In classical turbulence, the dissipation is important in the high wave number region, so that ST may show the characteristic behavior in this region, which will be investigated in future.
\bibitem{cs}  The velocity $c_{s}$ is the spin coherence length $\hbar / \sqrt{ |c_{1}| n M }$ divided by the time $\hbar / | c_{1} | n$. 
This time is characteristic of spin because $|c_{1}|n$ is the order of the spin interaction energy.
\end{thebibliography}
\end{document}